\documentclass[%
reprint,
superscriptaddress,
 amsmath,amssymb,
 aps,
]{revtex4-2}

\usepackage{newtxtext}
\usepackage{newtxmath}
\usepackage{natbib}
\usepackage{hyperref}
\usepackage{upgreek}
\usepackage{xcolor}
\usepackage{graphicx}
\usepackage{dcolumn}
\usepackage{bm}
\usepackage{orcidlink}

\begin{document}

\preprint{APS/123-QED}

\title{Elastic instability of wormlike micelle solution flow in serpentine channels}

\author{Emily Y. Chen,\orcidlink{0009-0004-6007-3460}}
\affiliation{%
 Department of Chemical and Biological Engineering, Princeton University, Princeton, NJ 08544, USA
}%

\author{Sujit S. Datta,\orcidlink{0000-0003-2400-1561}}%
 \email{Correspondence to ssdatta@caltech.edu}
\affiliation{%
 Division of Chemistry and Chemical Engineering, California Institute of Technology, Pasadena, CA 91125, USA
 }%
\affiliation{%
 Department of Chemical and Biological Engineering, Princeton University, Princeton, NJ 08544, USA
 }%

\date{\today}

\begin{abstract}
Wormlike micelle (WLM) solutions are abundant in energy, environmental, and industrial applications, which often rely on their flow through tortuous channels. How does the interplay between fluid rheology and channel geometry influence the flow behavior? Here, we address this question by experimentally visualizing and quantifying the flow of a semi-dilute WLM solution in millifluidic serpentine channels. At low flow rates, the base flow is steady and laminar, with strong asymmetry and wall slip. When the flow rate exceeds a critical threshold, the flow exhibits an elastic instability, producing spatially-heterogeneous, unsteady three-dimensional (3D) flow characterized by two notable features: (i) the formation and persistence of stagnant but strongly-fluctuating and multistable ``dead zones'' in channel bends, and (ii) intermittent 3D ``twists'' throughout the bulk flow. The geometry of these dead zones and twisting events can be rationalized by considering the minimization of local streamline curvature to reduce flow-generated elastic stresses. Altogether, our results shed new light into how the interplay between solution rheology and tortuous boundary geometry influences WLM flow behavior, with implications for predicting and controlling WLM flows in a broad range of complex environments.
\end{abstract}

\maketitle


\section{\label{sec:intro}Introduction}
Above a threshold concentration in solution, many surfactants self-assemble into long, flexible, polymer-like chains reaching microns in length. However, unlike polymers, these wormlike micelles (WLMs) constantly break and reform at thermal equilibrium and under flow. Hence, they are often termed ``living polymers''~\cite{larson_structure_1999,dreiss_chapter_2017}. When concentrated enough to be semi-dilute, WLM solutions are both strongly shear-thinning and highly elastic, and therefore, are used in a wide range of energy and industrial applications, such as subsurface proppant transport~\cite{chase_clear_1997,zana_giant_2007,jafari_nodoushan_wormlike_2019,du_environmentally_2023}, groundwater remediation~\cite{kwon_recent_2023,tripathy_biosurfactants_2024} , heating and cooling~\cite{gasljevic_saving_1993,shi_photoreversible_2013,kotenko_experimental_2019}, drag reduction~\cite{rose_drag_1989,gasljevic_two_2001,yang_viscoelastic_2002,rodrigues_drag_2020,mitishita_turbulent_2022}, lubrication~\cite{gupta_shear_2024}, drug manufacturing and delivery~\cite{chu_smart_2013,yu_intracellular_2013,lim_drugdependent_2022}, and as rheological modifiers in consumer products~\cite{zana_giant_2007,yang_viscoelastic_2002,dreiss_chapter_2017-1,choi_predicting_2020}. 

All of these applications involve the flow of WLM solutions through complex, confined spaces, such as winding pipes and porous rocks, where the boundary geometry introduces tortuosity to the flow. This tortuous flow in turn couples to the complex WLM solution rheology to produce unusual flow behaviors. For example, in Couette-type geometries, the homogeneous base flow can become unstable, separating into low and high shear bands that together maintain a constant shear stress with increasing shear rate~\cite{salmon_velocity_2003,dusek_shear-induced_2009,fielding_shear_2010,fardin_instabilities_2012,divoux_shear_2016,rassolov_kinetics_2017}---a form of shear localization. The high shear rate band can subsequently exhibit another instability that produces unsteady flow~\cite{becu_evidence_2007,fardin_elastic_2010,fardin_criterion_2011,fardin_instabilities_2012,beaumont_turbulent_2013,fielding_triggers_2016}, with pronounced spatiotemporal fluctuations similar to those exhibited by elastic polymer solutions in tortuous spaces~\cite{larson_purely_1990,larson_effect_1994,pakdel_elastic_1996,mckinley_rheological_1996,datta_perspectives_2022,browne_elastic_2021,browne2024harnessing}. This instability arises at low Reynolds number ($\mathrm{Re} \ll1$)---so the effects of inertia, which often causes flow instabilities, are negligible. Instead, it is generated by fluid elasticity, and is therefore often parameterized by the Weissenberg number, $\mathrm{Wi}$, which compares the relative strength of elastic and viscous stresses. This purely-elastic instability has also been documented for WLM solutions flowing in a range of microfluidic geometries, such as through rectangular channels~\cite{salipante_jetting_2017,haward_spatiotemporal_2014}, around sharp bends~\cite{hwang_flow_2017,zhang_flow_2018}, through contractions and expansions~\cite{salipante_flow_2018,matos_instabilities_2019}, in cavities~\cite{hillebrand_flow_2024}, and past a single obstacle~\cite{moss_flow_2010,haward_flow_2019,hopkins_purely_2020,hopkins_effect_2022,hopkins_upstream_2022,chen_flow_2004,wu_sphere_2018,wu_linear_2021} or two-dimensional (2D) array of obstacles~\cite{moss_flow_2010-1,de_flow_2018,haward_stagnation_2021}. However, such geometries have multiple confounding factors, such as mixed shear and extensional flow topology and the presence of stagnation points at solid surfaces, that can simultaneously influence the flow behavior. Thus, a clear understanding of how the coupling between tortuous boundary geometry and fluid rheology influences the complex flow behavior of WLM solutions is still lacking. 

Here, we simplify this problem by studying WLM solution flow at $\mathrm{Re}\ll1$ and $\mathrm{Wi}\sim1$ in serpentine channels comprising successive semi-circular half-loops. This microfluidic geometry enables us to isolate the role of fluid streamline curvature---which is known to be a key factor controlling purely-elastic flow instabilities since it generates elastic hoop stresses~\cite{pakdel_elastic_1996,mckinley_rheological_1996}---in shaping the unstable flow. Indeed, our work is inspired by prior studies of elastic instabilities in polymer solutions~\cite{groisman_efficient_2001,burghelea_chaotic_2004,zilz_geometric_2012,casanellas_stabilizing_2016,soulies_characterisation_2017,shakeri_characterizing_2022}; these provide useful intuition, but are not directly reflective of the flow behavior of WLM solutions, which have fundamentally different microstructures and rheological properties. We find that the base steady, laminar flow exhibits strong asymmetry and wall slip at the lowest flow rates tested due to the unique rheology of the WLM solution, which enables it to support shear localization. Above a critical $\mathrm{Wi}=\mathrm{Wi}_c$, the flow becomes unstable, producing spatially-heterogeneous unsteady three-dimensional (3D) flow. This unsteady flow has two notable features: (i) persistent, fluctuating dead zones, which form due to the ability of the WLM solution to support shear localization, and (ii) slow ``twisting'' events in 3D. Our characterization of these unusual flow behaviors suggests that when WLM solutions are forced to flow in confined, tortuous spaces, they develop pathways of least resistance that minimize local streamline curvature. Altogether, our work provides a key first step toward making sense of complex fluid flows in complex spaces, potentially enabling their prediction and control in a broad range of applications.\\

\section{Methods\label{sec:methods}}
\subsection{\label{sec:fluid} Test fluid and preparation}
We use a well-characterized semi-dilute, entangled WLM solution composed of $100$~mM cetylpyridinium chloride (CPyCl; Sigma-Aldrich) and $60$~mM sodium salicylate (NaSal; Sigma-Aldrich) in ultrapure Millipore water~\cite{rehage_rheological_1988,mair_observation_1996,oelschlaeger_linear--branched_2009,haward_stagnation_2012,haward_extensional_2012,gaudino_adding_2015,salipante_flow_2018,matos_instabilities_2019,haward_stagnation_2021,hopkins_upstream_2022,hopkins_effect_2022,hillebrand_flow_2024}. We prepare the solution by first dissolving the CPyCl, which is a cationic surfactant, and then adding the NaSal, an aromatic organic salt, and stirring the solution for $24$~h. The addition of NaSal induces the elongation of micelles due to electrostatic screening of the charged headgroups and steric interaction with the aromatic ring of the organic salt~\cite{gaudino_adding_2015}. We let the resulting WLM solution rest for $>48$~h to allow air bubbles to disappear before use. We also add 1~$\upmu$m diameter fluorescent, amine-functionalized polystyrene tracer particles (Invitrogen; yellow-green) to the solution at 20~ppm for flow visualization and particle image velocimetry (PIV).

\subsection{\label{sec:rheo} Shear rheology}
\begin{figure}
\centering
  \includegraphics{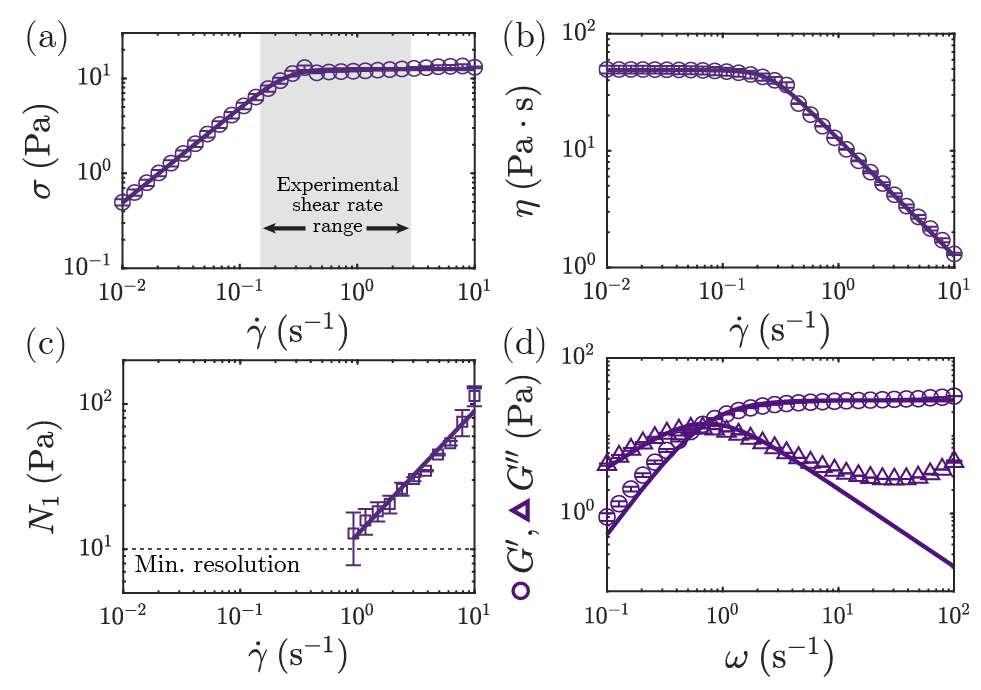}
  \caption{\textbf{Shear rheology of 100/60 mM CPyCl/NaSal WLM solution.} (a) Shear stress $\sigma$ and (b) shear viscosity $\eta$ as a function of imposed shear rate $\dot{\gamma}$, exhibiting a shear stress plateau characteristic of shear banding. Solid lines correspond to a Carreau-Yasuda fit. (c) First normal stress difference $N_1$ as a function of imposed shear rate $\dot{\gamma}$, with power law fit (solid line). (d) Storage $G'$ and loss $G''$ moduli as a function of angular frequency $\omega$ under small amplitude oscillatory shear in the linear viscoelastic regime at 1\% strain amplitude. Solid curves are fits to the single-mode Maxwell model with a characteristic relaxation time $\lambda_M = 1.4$~s. Error bars correspond to the standard deviation over 3 replicate measurements. }
  \label{fig:rheo}
\end{figure}

To characterize the shear rheology of the WLM solution, we use a stress-controlled Anton Paar MCR501 rheometer fitted with a truncated cone-plate geometry (CP50-2: 50 mm diameter, 2$^\circ$, 53 $\upmu$m) at a controlled temperature of 25$^\circ$C. We measure steady-state flow curves at fixed values of the shear rate $\dot{\gamma}$, ramping up and down between $0.01$ and $10\ \mathrm{s^{-1}}$; shear rates exceeding $10 \ \mathrm{s^{-1}}$ result in fluid ejection due to an elastic instability. The results, averaged over three replicate measurements, are shown in Figure~\ref{fig:rheo}(a)-(b). As shown in Fig.~\ref{fig:rheo}(a), the shear stress $\sigma$ increases approximately linearly with shear rate up to $\dot{\gamma}_0\approx0.3~\mathrm{s^{-1}}$, and then plateaus at a value $\sigma^*=13$ Pa, indicative of shear banding~\cite{dusek_shear-induced_2009,fielding_shear_2010,fardin_instabilities_2012,divoux_shear_2016}. As we show later on, this property of the WLM solution has an important consequence for its flow in serpentine channels: it enables them to support shear localization, i.e., separate into low and high shear regions that together maintain an effectively constant bulk shear stress. As shown in Fig.~\ref{fig:rheo}(b), the corresponding dynamic viscosity $\eta(\dot{\gamma})=\sigma(\dot{\gamma})/\dot{\gamma}$ is nearly constant at $\eta_0 = 49$~Pa$\cdot$s at low shear rates, and then continually decreases (shear thinning) with increasing $\dot{\gamma}>\dot{\gamma}_0$. This behavior is described well by the Carreau-Yasuda model~\cite{macosko_rheology_1994}, as shown by the solid lines: $\eta(\dot{\gamma})=\eta_\infty+(\eta_0-\eta_\infty)(1+(\dot{\gamma}/\dot{\gamma}_0)^a)^{(n-1)/a} $, where $a=3.8$ describes the transition to the shear-thinning regime, the power law index $n\approx0$, and we set $\eta_\infty$ equal to the solvent viscosity of $1$~mPa$\cdot$s given the absence of a measurable high shear rate Newtonian plateau.

To characterize the solution elasticity, we also measure the first normal stress difference $N_1$ and linear elastic storage and loss moduli $G'$ and $G''$, respectively. As shown in Fig.~\ref{fig:rheo}(c), $N_1$ increases with shear rate as a power law (solid curve), $N_1(\dot{\gamma})=K_{N_1}\dot{\gamma}^{n_{N_1}}$, with $K_{N_1} = 12.7 \ \mathrm{Pa\cdot s}^{n_{N_1}}$ and power law index $n_{N_1} = 0.85$. The instrument resolution due to normal force sensitivity is roughly  $N_1 \approx 10$ Pa; thus, estimates of $N_1$ using the power law model at shear rates below $\dot{\gamma}\approx 1 \ \mathrm{s^{-1}}$ are extrapolations of this power law. As shown by the curves in Fig.~\ref{fig:rheo}(d), $G'$ and $G''$ determined from small-amplitude oscillation over a range of angular frequencies $\omega$ are well described by a single-mode Maxwell model: $G'(\omega)=\frac{G_0\lambda_M^2\omega^2}{1+\lambda_M^2\omega^2}$ and $G''(\omega) = \frac{G_0 \lambda_M\omega}{1+\lambda_M^2\omega^2}$, where $G_0 = 29$~Pa is the plateau modulus and $\lambda_M = 1.4$~s is the Maxwell relaxation time. We thereby estimate the mesh size of the entangled micellar network as $\xi=\left(\frac{k_B T}{G_0}\right)^{\frac{1}{3}}= 52$ nm~\cite{larson_lengths_2012}. Based on prior estimates of the persistence length for the same WLM solution ($l_p\approx28$ nm~\cite{oelschlaeger_linear--branched_2009,gaudino_adding_2015}), we estimate the average micelle contour length to be $\sim830 \ \upmu$m~\cite{granek_stress_1992,gaudino_adding_2015}. The rheology measurements also enable estimates of the micelle breakage and reptation time scales $\tau_{\mathrm{break}} = 0.49$~s and $\tau_{\mathrm{rept}} = 3.5$~s, respectively, using the framework of Turner and Cates~\cite{turner_linear_1991}. Their ratio $\frac{\tau_{\mathrm{break}}}{\tau_{\mathrm{rept}}} = 0.14 \ll 1$, confirming that the micelles are in the fast-breaking limit where the fluid is well-described by the idealized Maxwell model~\cite{turner_linear_1991}. 

\subsection{\label{sec:device} Device architecture and fabrication} 
\begin{figure}
\centering
  \includegraphics{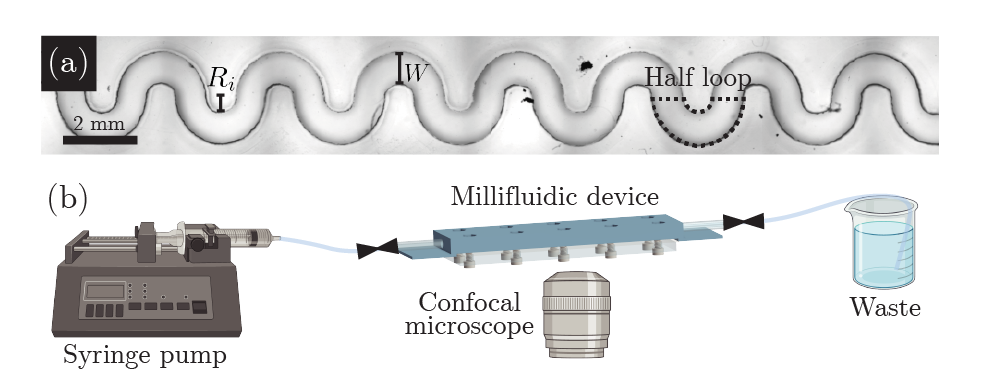}
  \caption{\textbf{Serpentine channel geometry and experimental setup. }(a) Brightfield optical image of the serpentine channel device comprised of 19 semi-circular half-loops with inner radius $R_i$ and channel width $W$. (b) Schematic of experimental setup for flow visualization.}
  \label{fig:expt}
\end{figure}

The serpentine channels used for the flow experiments comprise 19 semi-circular half-loops with inner radius $R_i$, fixed channel width $W = 1$~mm, outer radius $R_o=R_i+W$, fixed channel height $H=2$~mm, and fixed aspect ratio $H/W=2$; an optical top-down image of a representative example is shown in Fig.~\ref{fig:expt}(a). We choose a large channel aspect ratio to ensure that the dominant velocity gradients arise in the flow imaging plane. All the quantitative results presented in this paper are for channels with $R_i=300~\upmu$m; however, as shown in SI Movies 3 and 4, we observe similar behavior in channels with $R_i=500$ and  $1000~\upmu$m.

We design the body of each channel [dark blue in Fig.~\ref{fig:expt}(b)] using CAD software (Onshape) and 3D-print it with a clear methacrylate-based polymeric resin (FLGPCL04) using a FormLabs Form 3 stereolithography printer. We assemble each millifluidic device by screwing together this 3D-printed channel body and a laser-cut transparent acrylic top sheet (McMaster-Carr) [gray in Fig.~\ref{fig:expt}(b)], with a $\sim1$~mm-thick layer of polydimethylsiloxane (PDMS; Dow SYLGARD 184, 8.5:1.5 base:curing agent by weight) sandwiched in between to act as a gasket. We then glue flexible Tygon tubing (McMaster-Carr) in the inlet and outlet using a 2-part watertight epoxy (JB MarineWeld).

\subsection{\label{sec:protocol} Flow imaging}
Before each flow experiment, we flush the device to be used with ultrapure Millipore water to remove any residual debris or air bubbles. We then fully saturate the channel with the WLM solution to be tested at a flow rate of $1$~mL/h for at least $3$~h using a syringe pump (Harvard Apparatus PHD 2000). The device is affixed to the stage of an inverted Nikon
A1R+ laser scanning confocal fluorescence microscope, ensuring that the syringe pump, device, and outlet
waste jar are at the same height to avoid hydrostatic pressure
differences, as shown in Fig.~\ref{fig:expt}(b).

In each experiment, we impose a stepwise ramp of increasing flow rate $Q$ from $0.5$ to $6$~mL/h, allowing the flow to equilibrate at each flow rate for $30$~min before commencing imaging. For each flow rate tested, we use a 4$\times$ objective lens to acquire $2$~min long ($\sim86 \lambda_M$) movies at $30$~frames per second focused on a horizontal optical plane,  $37~\upmu$m in thickness, centered at the mid-channel height to avoid boundary effects induced by the top and bottom channel walls. Each movie is taken at each successive half-loop along the length of the serpentine array, with a $3167\times3167~\upmu$m$^2$ field of view at a resolution of 6~$\upmu$m/pixel. The fluorescent tracer particles seeded in the fluid are excited by a $488$~nm laser and their emission is detected using a $500-550$~nm sensor. 

To characterize the flow behavior, we define a nominal shear rate using the channel half-width $W/2$ as the characteristic length scale: $\dot{\gamma}\equiv\frac{Q/\left(HW\right)}{W/2}$. Our experiments probe the range $0.14 \leq \dot{\gamma} \leq 1.7 ~\mathrm{s^{-1}}$; as shown by the gray region in Fig.~\ref{fig:rheo}(a), this range corresponds to strongly shear-thinning conditions for the WLM solution. We thereby define the Weissenberg number $\mathrm{Wi}$, which compares the relative strength of elastic and viscous stresses, as $\mathrm{Wi}\equiv\frac{N_1(\dot{\gamma})}{2\eta(\dot{\gamma})\dot{\gamma}}$, where $N_1(\dot{\gamma})$ and $\eta(\dot{\gamma})$ are given by the power law and Carreau-Yasuda fits to the bulk shear rheology, respectively. Our experiments probe the range $0.17<\mathrm{Wi}<0.78$, and the onset of shear banding arises at $\mathrm{Wi}_0=\mathrm{Wi}(\dot{\gamma}_0)=0.19$. For flow conditions exceeding $\mathrm{Wi}=0.78$, we are unable to obtain PIV measurements for quantitative analysis.

\begin{figure}
\centering
  \includegraphics{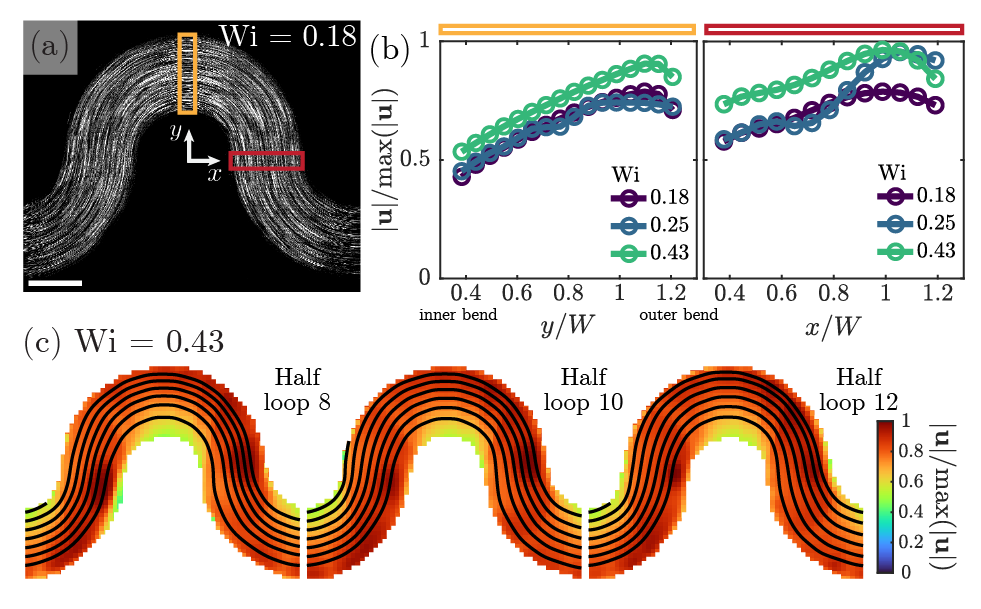}
  \caption{\textbf{Base steady laminar flow exhibits strong shear-thinning and slip at surfaces.} (a) Pathline image of flow in the $R_i = 300\ \upmu$m device at $\mathrm{Wi} = 0.18 < \mathrm{Wi}_c$, generated using a moving average of fluorescence intensity over 60 frames. Scale bar is $500\ \upmu$m. (b) Time-averaged (over $86\lambda$) velocity profiles, normalized by the maximum velocity, at the apex (yellow rectangle) and end (red rectangle) of a half-loop. All velocity profiles exhibit highly asymmetric plug flow and indicate the presence of a substantial slip velocity at surfaces. The PIV grid resolution used is 0.05$W$. (c) Streamline visualization and color maps of the normalized velocity magnitudes at selected half-loop locations further reflect strong plug flow and asymmetry in the base laminar flow state.}
  \label{fig:base}
\end{figure}

\begin{figure*}
\centering
  \includegraphics{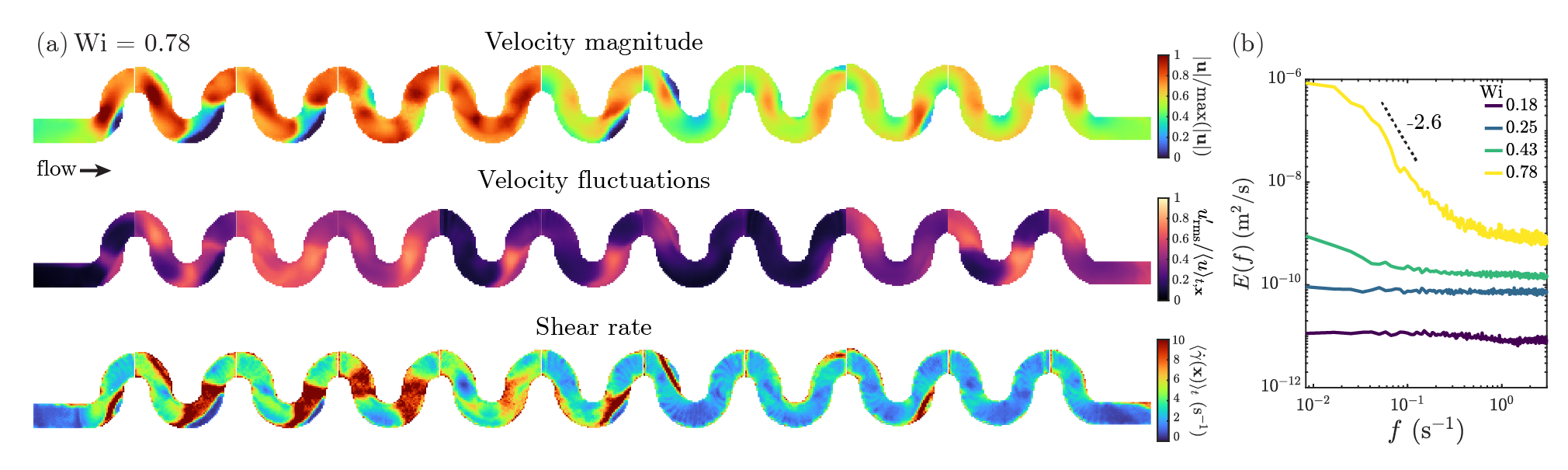}
  \caption{\textbf{Unsteady 3D flow arising from an elastic instability.} (a) Instantaneous color maps in the unsteady flow state showing the (top) normalized velocity magnitude $|\mathbf{u}|/\mathrm{max}(|\mathbf{u}|)$, (middle) root-mean-square (rms) velocity magnitude fluctuations $u_{\mathrm{rms}}'/\langle u \rangle_{t,\mathbf{x}}$, and (bottom) shear rate $\dot{\gamma}(x)$ in the $R_i = 300\ \upmu$m channel at $\mathrm{Wi} = 0.78 > \mathrm{Wi}_c$. The top and bottom rows show near-instantaneous flow fields that are time-averaged over a duration of 1$\lambda$ to reduce noise. The middle row shows rms velocity fluctuations over a duration of 86$\lambda$, highlighting the spatial heterogeneity and temporal persistence of flow fluctuations. All color maps comprise stitched fields-of-view obtained sequentially during flow imaging, giving rise to the visible discontinuities between adjacent half-loops. (b) Power spectral densities of the magnitude of velocity fluctuations indicate that above the onset of the elastic instability, more energy is dissipated in fluctuations across a broad range of frequencies, with no characteristic time scale. 
  }
  \label{fig:onset}
\end{figure*}

To visualize the WLM solution flow, we generate fluid pathline images and movies using grouped projections of mean pixel intensity over successive frames. Example movies are shown in SI Movies 1-4. Quantitative analyses of the flow field are done by directly measuring the time-resolved 2D velocity fields, $\mathbf{u}$, using particle image velocimetry (PIVlab~\cite{thielicke_pivlab_2014}) with a grid resolution of $50~\upmu$m. We characterize the flow in each channel using three metrics: the normalized velocity magnitude $|\mathbf{u}|/\mathrm{max}(|\mathbf{u}|)$, root mean square velocity fluctuations $u_{\mathrm{rms}}'/\langle u \rangle_{t,\mathbf{x}}$, and shear rate $\dot{\gamma}(\textbf{x})=\sqrt{2\mathbf{D}:\mathbf{D}}$, where $\textbf{x}$ and $t$ represent the 2D position in the horizontal optical plane and time, respectively, and $\mathbf{D}=\frac{1}{2}(\nabla \mathbf{u}+\nabla \mathbf{u}^\intercal)$ is the rate-of-strain tensor computed from the measured velocity field.\\

\section{Results and discussion\label{sec:results}}

\subsection{\label{subsec:base} Base steady laminar flow}
We first characterize the steady base flow of the WLM solution in the serpentine channels at low $\mathrm{Wi}$. A representative pathline image of the laminar steady flow at $\mathrm{Wi}=0.18$ is shown in Fig.~\ref{fig:base}(a), and the velocity profiles at the apex and exit of a half-loop are shown for the regions outlined in yellow and red in Fig.~\ref{fig:base}(b), respectively; the corresponding SI Movie 1 shows that the pathlines remain unchanging over time. Notably, at the apex of the bend, the velocity profile is highly asymmetric and skewed toward the outer bend. While the curvature of the channel walls is known to induce slight asymmetry in Newtonian flows~\cite{zilz_geometric_2012,soulies_characterisation_2017,ducloue_secondary_2019,shakeri_characterizing_2022}, in this case the velocity profiles are starkly different. In particular, as shown in Fig.~\ref{fig:base}(b), the flow is plug-like, unlike the typical parabolic profile associated with Poiseuille flow, because of the shear thinning of the WLM solution. Additionally, near the channel walls, the velocity does not decrease to zero---indicating strong slip caused by shear localization at the fluid-solid boundary, as has been reported for WLM flows in other geometries~\cite{mair_observation_1996,fardin_instabilities_2012,haward_extensional_2012,gupta_shear_2024}. Interestingly, we observe hints of shear inhomogeneity when $\mathrm{Wi}$ is increased slightly to $0.25$, as shown by the inflection in the blue profile shown in the right-hand plot of Fig.~\ref{fig:base}(b). This effect diminishes upon further increasing $\mathrm{Wi}$, as shown by the green profile in Fig.~\ref{fig:base}(b), suggesting that the flow re-homogenizes across the channel width. A representative image showing the velocity magnitudes and flow streamlines across multiple half-loops is shown in Fig.~\ref{fig:base}(c) for this case of $\mathrm{Wi}=0.43$, at which the flow remains laminar and steady.

\subsection{\label{subsec:EI} Onset of 3D unsteady flow}
Further increasing $\mathrm{Wi}$ gives rise to an elastic flow instability where this base steady flow becomes unstable to 3D unsteady flow. An example is shown in Fig.~\ref{fig:onset}(a); for this channel geometry, the onset of the instability---determined as when the rms velocity fluctuations exceed a threshold value of $0.2$ set by the noise threshold of the PIV measurements---occurs at a critical value $\mathrm{Wi}=\mathrm{Wi}_c\approx0.7$. Such purely-elastic instabilities are typically generated through the coupling of elastic normal stresses and streamline curvature, which gives rise to destabilizing hoop stresses~\cite{larson_purely_1990,pakdel_elastic_1996,mckinley_rheological_1996,datta_perspectives_2022}. This coupling can also be described using a dimensionless parameter developed by Pakdel and McKinley~\cite{pakdel_elastic_1996,mckinley_rheological_1996}: $\mathrm{M}\equiv\sqrt{2\mathrm{Wi}\cdot\frac{\lambda_M U}{R_i}}$, where $U=Q/\left(HW\right)$ is the average flow velocity. In our experiments, the critical value of $\mathrm{M}=\mathrm{M}_c\approx2.5$, in good agreement with prior measurements across a broad range of fluids and geometries~\cite{datta_perspectives_2022}.

As shown by the maps in Fig.~\ref{fig:onset}(a), all three flow metrics exhibit pronounced heterogeneity throughout the channel. To further characterize the flow dynamics, we examine the temporal power spectral density (PSD) of the magnitude of velocity fluctuations across different frequencies $f$: $E(f)=\frac{1}{LF_s}|\mathrm{FFT}(u'(t))|^2$, where $L$ is the duration of the signal, $F_s$ is the sampling rate, and $\mathrm{FFT}$ denotes a fast Fourier transform. Our results, spatially averaged over all PIV voxels, are shown in Fig.~\ref{fig:onset}(b). Above the onset of the elastic instability, the PSD exhibits a broad power law decay $E(f)\sim f^{-\alpha}$, with $\alpha \approx 2.6$ at $\mathrm{Wi} = 0.78$, reflecting energy dissipation by flow fluctuations across a broad range of time scales---consistent with previous characterizations of polymeric and WLM elastic instabilities in other confined geometries~\cite{groisman_elastic_2004,fardin_elastic_2010,fardin_instabilities_2012,soulies_characterisation_2017,salipante_flow_2018,matos_instabilities_2019,browne_elastic_2021,carlson_volumetric_2022,shakeri_characterizing_2022,chen_stagnation_2024}. The absence of distinct peaks in the PSD indicates that the unsteady flow is aperiodic with no characteristic time scales.\\

\begin{figure}
\centering
\includegraphics{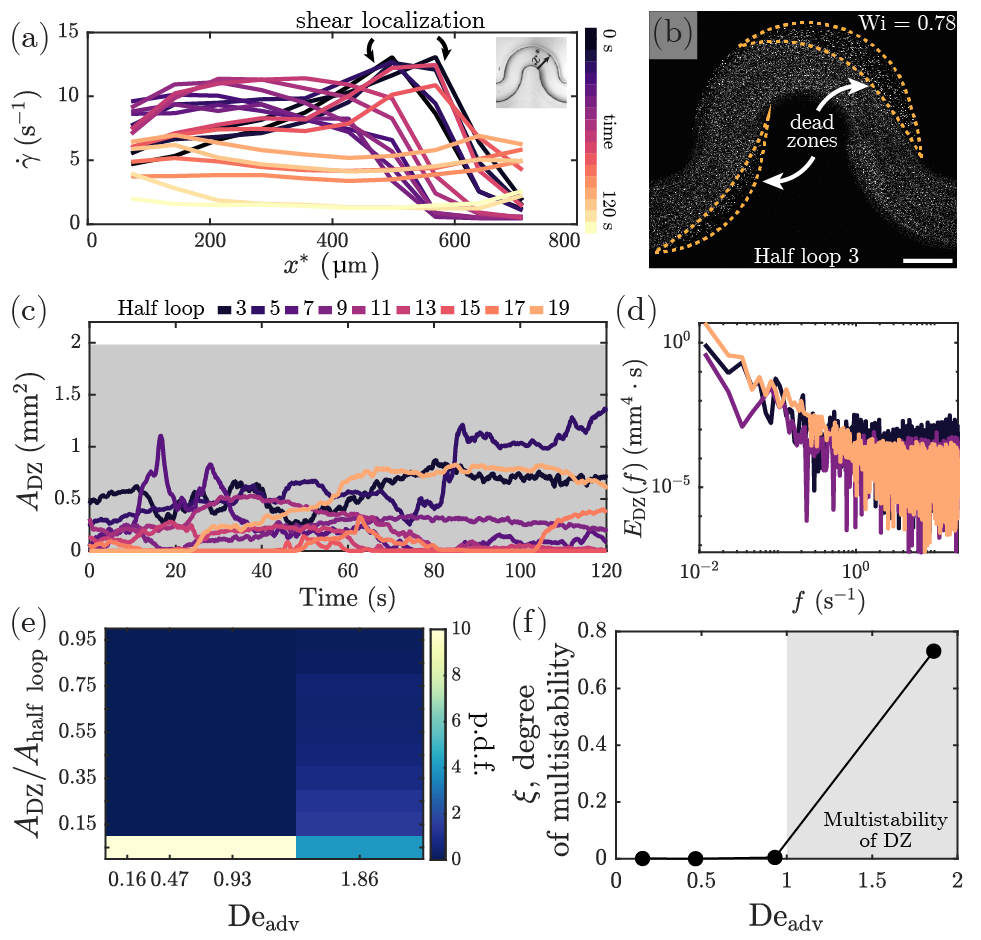}
  \caption{\textbf{Characterization of dead zone formation, dynamics, and size.} (a) Temporal evolution of shear rate profiles along the channel width (location shown in inset). Successive curves are plotted at time intervals of $6.7$~s. At early times ($0<t<90$~s), a large peak in shear rate separates the dead zone ($\dot{\gamma} \approx 0 \ \mathrm{s^{-1}}$) from the bulk flow, while at later times the dead zone is washed away and the shear rate profile re-homogenizes. (b) Example instantaneous pathline image of the unsteady flow state in a selected half-loop. Yellow dashed lines demarcate example dead zone regions. Scale bar shows $500 \ \upmu$m. (c) Temporal variation of dead zone size $A_{\mathrm{DZ}}$ in selected half-loops shows considerable fluctuations. The shaded gray region denotes the range of theoretically-predicted dead zone areas from considering minimization of local streamline curvature. (d) Power spectral densities of dead zone size fluctuations indicate no characteristic time scale. (e) Probability density distribution of normalized dead zone areas $A_{\mathrm{DZ}}/A_{\mathrm{half\ loop}}$ with increasing advective Deborah number $\mathrm{De_{adv}}$ shows a broad increase in dead zone sizes for $\mathrm{De_{adv}}>1$. (f) Degree of multistability $\xi$ describing the range of observed dead zone sizes at a given flow condition shows onset of multistable dead zone behavior above $\mathrm{De_{adv}} = 1$. All data in (a)-(d) are for $\mathrm{Wi}=0.78$. 
  }
  \label{fig:DZ}
\end{figure}

\subsubsection{\label{DZsec} Dead zone formation, dynamics, and size are shaped by fluid rheology.} 
Close inspection of the unstable flow shown in Fig.~\ref{fig:onset}(a) reveals a fascinating consequence of the WLM solution's ability to support shear banding: As shown by the dark blue regions in the top row of Fig.~\ref{fig:onset}(a), non-flowing ``dead zones'' (identified by $|\mathbf{u}|/\mathrm{max}(|\mathbf{u}|) < 0.05$) form and persist in the downstream portion of some of the half-loops. The corresponding SI Movie 2 shows these dynamics using pathline imaging. While they are reminiscent of corner eddies reported for viscoelastic WLM solution flows around sharp bends~\cite{hwang_flow_2017,zhang_flow_2018}, the dead zones formed in our serpentine channels are truly near-stagnant regions with minimal flow, whereas eddies are characterized by non-zero circulating flow. This feature is illustrated by the shear rate profiles across the channel width (inset) in Fig.~\ref{fig:DZ}(a), and an example instantaneous pathline image with two dead zones demarcated by the yellow dashed outlines is shown in Fig.~\ref{fig:DZ}(b). As shown in the dark-colored profiles in Fig.~\ref{fig:DZ}(a), a clear peak of high shear rate separates the dead zone ($\dot{\gamma}\approx0 \ \mathrm{s^{-1}}$) and the flowing regions, generating a strong shear rate gradient at the dead zone boundary; by contrast, if the fluid could not support shear localization, dead zones would not form, as previously documented for non-shear banding elastic polymer solutions~\cite{zilz_geometric_2012,casanellas_stabilizing_2016,soulies_characterisation_2017,shakeri_characterizing_2022}.

  To characterize their dynamics, we track variations in the dead zone areas ($A_{\mathrm{DZ}}$) over time. Examples for selected half loops along the length of the channel are shown by the traces in Fig.~\ref{fig:DZ}(c). Notably, the dead zone sizes fluctuate dramatically over time, often persisting over durations much longer than the characteristic relaxation time $\lambda_M=1.4$~s. These fluctuations do not have any clear periodicity; instead, they mirror the spectrum of bulk velocity fluctuations shown in Fig.~\ref{fig:onset}(b), as reflected by the broad decay in the power spectral density $E_{\mathrm{DZ}}(f)$ in Fig.~\ref{fig:DZ}(d)---indicating that dead zone fluctuations are established by the unstable fluctuations in the freely-flowing fluid outside the dead zone. Despite these fluctuations, however, all dead zones are bounded by a maximal size $A_{\mathrm{DZ}}^{\mathrm{max}}$, indicated by the gray region in Fig.~\ref{fig:DZ}(c).

Intriguingly, for a given unstable flow rate and at a given time, not all half-loops contain dead zones---as shown in Figs.~\ref{fig:onset}(a) and ~\ref{fig:DZ}(c). Instead, they exhibit multistable behavior: some half-loops do not have dead zones (the dead zone-free state), while other half-loops have persistent dead zones of varying sizes (the dead zone-containing state). As time progresses, a given half-loop can randomly switch from the dead zone-free state to the dead zone-containing state and vice versa; an example is shown by the different traces in Fig.~\ref{fig:DZ}(a), in which a dead zone eventually washes away, as well as those in Fig.~\ref{fig:DZ}(c). This behavior is reminiscent of the multistability exhibited by eddies during the unstable flow of elastic polymer solutions in pore constriction arrays~\cite{browne_bistability_2020,kumar_numerical_2021,chen_influence_2025}, in which multistability arises when flow-induced elastic stresses do not have sufficient time to relax as they are advected between constrictions. Hence, following this prior work, we parameterize the onset of the multistability observed for WLM solutions in serpentine channels using a streamwise Deborah number comparing the stress relaxation time of the WLM solution to the time required for the fluid to be advected between adjacent half-loops: $\mathrm{De_{adv}}\equiv\frac{\lambda_M}{t_{\mathrm{adv}}}$ where $t_{\mathrm{adv}}=HA_{\mathrm{half \ loop}}/Q$ and $A_{\mathrm{half \ loop}}$ is the half-loop area. Consistent with this prior work, we also find that multistability arises when $\mathrm{De_{adv}}\gtrsim1$. For example, Fig.~\ref{fig:DZ}(e) shows a color map of the probability density function (p.d.f.) of the dead zone areas aggregated over the different half-loops, normalized by $A_{\mathrm{half \ loop}}$, for the different $\mathrm{De_{adv}}$ tested---showing that this p.d.f. abruptly begins to broaden, a hallmark of multistability, when $\mathrm{De_{adv}}\gtrsim1$. Furthermore, also following prior work, we define the degree of multistability $\xi \equiv \frac{\mathrm{max}(A_{\mathrm{DZ}})-\mathrm{min}(A_{\mathrm{DZ}})}{A_{\mathrm{half \ loop}}}$, where $\xi = 0$ indicates no multistable behavior and $\xi=1$ describes the maximum possible extent of multistable behavior. As shown in Fig.~\ref{fig:DZ}(f), we again see an abrupt increase in the degree of multistability $\xi$ when $\mathrm{De_{adv}}\gtrsim1$. Taken together, these results demonstrate that fluid elasticity plays a key role in determining dead zone formation. 

Motivated by these findings, as well as previous work describing the shapes of eddies formed in elastic polymer solutions entering contractions~\cite{batchelor_stress_1971,boger_viscoelastic_1987,mongruel_extensional_1995,mongruel_axisymmetric_2003,rodd_role_2007}, we conjecture that considering the elastic stresses generated during WLM solution flow in a serpentine channel can help predict the maximal dead zone size $A_{\mathrm{DZ}}^{\mathrm{max}}$. In particular, given that these elastic hoop stresses are generated by streamline curvature~\cite{pakdel_elastic_1996,mckinley_rheological_1996}, we expect that the interface between a dead zone and the surrounding freely-flowing region forms to minimize local streamline curvature. This idea can be formulated using a simple geometric description (detailed in the Appendix): for a given half-loop, we expect that the largest dead zone that can form in a half-loop is bounded by a straight line tangent to the inner bend of the loop, as shown in Fig.~\ref{fig:app}. This prediction, shown by the gray shaded region in Fig.~\ref{fig:DZ}(c), agrees reasonably well with the experimental measurements as an upper bound $A_{\mathrm{DZ}}^{\mathrm{max}}$ to the individual traces of $A_{\mathrm{DZ}}$.

\begin{figure}
\centering  \includegraphics{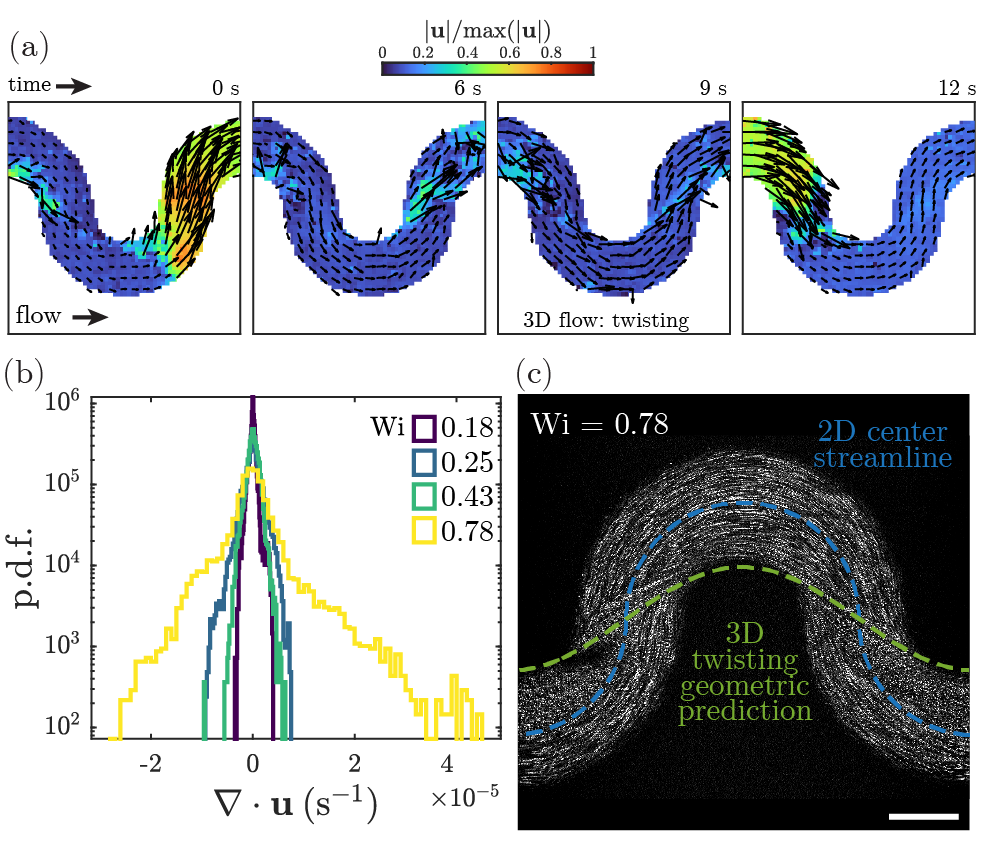}
  \caption{\textbf{3D twisting flows intermittently arise in the unsteady flow.} (a) Snapshots of the velocity field (arrows) superimposed on a color map of normalized velocity magnitude at different times, showing a representative twisting event. (b) Probability density function of the divergence of the measured 2D flow field suggests an increase in 3D flow above the onset of instability ($\mathrm{Wi}>\mathrm{Wi}_c\approx0.7$). (c) Twisting reduces the hydrodynamic tortuosity of fluid streamlines: the blue dashed curve is the center semi-circular streamline in the base flow, and the green dashed curve is the projected 2D sinusoidal path enabled by 3D twisting. Pathline images are generated using a moving average of fluorescence intensity over 10 frames. Scale bar is $500~\upmu$m. Data in (a) and (c) are for $\mathrm{Wi}=0.78$.}
  \label{fig:twist}
\end{figure}

\subsubsection{\label{Twistsec} 3D twisting flow events reduce local hydrodynamic tortuosity.}
SI Movie 2 also reveals another unusual feature of the unstable WLM solution flow: 3D flow inversion events characterized by lower velocity ``twists'' in the imaging plane. Additional examples are shown in SI Movie 4. Figure~\ref{fig:twist}(a) presents snapshots taken at different times of the fluid velocity field (arrows, with the normalized velocity magnitude shown by the color map) corresponding to one such twisting event. The initial faster-flowing, primarily 2D flow (bright green/yellow/red region in the first panel) is overtaken by a slower-moving flow (dark blue region) that twists in the $z$-direction, as reflected by a curvy ``disclination'' of the velocity vectors. Fig.~\ref{fig:twist}(b) confirms the presence of 3D flow in this unsteady flow state by plotting the probability density function of the divergence of the 2D velocity field, $\nabla \cdot \mathbf{u}=\frac{\partial u}{\partial x}+\frac{\partial v}{\partial y}$, using the in-plane velocity components. Under steady flow conditions ($\mathrm{Wi=0.18-0.43}$), the 2D divergence is close to zero, with slight deviations due to small weakly-3D secondary flows generated by the inversion of curvature at corners and bends~\cite{rusconi_secondary_2011,poole_viscoelastic_2013,kim_filaments_2014,ducloue_secondary_2019}; however, in the unstable flow state ($\mathrm{Wi=0.78}$), an appreciable fraction of the 2D divergence becomes non-zero, reflecting flow in the third dimension.

Despite their complexity, we attempt to rationalize the basic geometric properties of these 3D inversion events by again considering minimization of the local streamline curvature---which here manifests as a reduction of the 2D hydrodynamic tortuosity, $\tau$. In the base laminar flow, the streamlines can be approximated as a semi-circular path shown by the blue dashed line in Fig.~\ref{fig:twist}(c); the associated tortuosity, given by the ratio of the total path length to the horizontal streamwise distance traveled, is then $\tau=\frac{\pi}{2}$. Guided by the visualization in SI Movie 4 and Fig.~\ref{fig:twist}(a), we approximate the ``disclination'' boundary of the twisting flow as a periodic sinusoidal function with amplitude $R_i$ and period $2(R_i+R_o)$. As shown by the green dashed line in Fig.~\ref{fig:twist}(c), this approximation is in good agreement with the experimental observations. The resulting 2D hydrodynamic tortuosity is then given by $\tau=\frac{1}{2R_i+2R_o} \int_0^{2R_i+2R_o} \sqrt{1+(\frac{\mathrm{d}y}{\mathrm{d}x})^2}\mathrm{d}x$, where $x$ and $y$ represent the streamwise and transverse position coordinates. For this channel, this calculation then yields a tortuosity of $1.08$, representing a $31\%$ reduction from the laminar base case.\\

\section{Conclusions}
In this study, we investigated the flow behavior of a highly elastic, shear-thinning, semi-dilute WLM solution in serpentine channels at low Reynolds number ($\mathrm{Re} \ll1$) and moderate Weissenberg numbers ($\mathrm{Wi} \sim1$). The serpentine microchannels are composed of successive semicircular half-loops, thereby enabling us to experimentally isolate the influence of streamline curvature---a fundamental feature of many real-world flows---on the flow behavior without being confounded by other geometric complexities. Our flow visualization experiments revealed three key phenomena:
\begin{enumerate}
    \item At low $\mathrm{Wi}$, the base flow is steady and laminar but exhibits spatial asymmetry with wall slip, reflecting the shear-thinning and shear banding properties of the WLM solution. Above a critical $\mathrm{Wi}=\mathrm{Wi}_c\approx0.7$, the flow undergoes an elastic instability and transitions to a 3D unsteady flow state characterized by pronounced spatiotemporal velocity fluctuations. This transition occurs at a critical Pakdel-McKinley parameter value of $\mathrm{M}_c\approx2.5$, consistent with other measurements across different viscoelastic fluids and geometries.
    \item Alongside this unstable bulk flow, dead zones of stagnant fluid form in the downstream portion of half-loops---reflecting the ability of the WLM solution to support shear localization, complementing reports of dead zone formation for other types of complex fluids~\cite{coussot_yield_2014,waisbord_anomalous_2019,abdelgawad_interplay_2024,kawale_polymer_2017,ichikawa_viscoelastic_2022}. Due to coupling to the velocity fluctuations in the bulk flow, these dead zones fluctuate in their size; however, they are bounded by a maximal size that minimizes the fluid streamline curvature, and therefore the generation of elastic stresses. Dead zones also exhibit multistable behavior---forming and persisting in some half-loops, not forming in other half-loops, and randomly switching between these two states. This multistability arises when the advective Deborah number $\mathrm{De}_{\mathrm{adv}}\gtrsim 1$, indicating that it occurs because elastic stresses generated by fluid flow do not have sufficient time to relax between consecutive half-loops. This finding expands on previous studies of multistability, which were restricted to elastic polymer solutions flowing through pore constriction arrays~\cite{browne_bistability_2020,kumar_numerical_2021,chen_influence_2025}, to a broader range of fluids and flow geometries.
    \item The unstable flow state also features intermittent, 3D ``twisting'' velocity inversion events amid the spatiotemporally-fluctuating bulk flow. These twisting events reduce the hydrodynamic tortuosity compared to the base flow state, and their geometric structure can also be rationalized as minimizing the fluid streamline curvature, and therefore the generation of elastic stresses.    
\end{enumerate}
These findings highlight how the shear banding and elastic properties of WLM solutions give rise to unusual flow behaviors in tortuous channels, expanding current understanding of complex fluid flows in complex geometries. Further investigating the physics underlying the formation of the dead zones and 3D twisting events revealed by our experiments will be a useful direction for future work. 

Several extensions of our study offer opportunities for future research. While our imaging focused on a single optical plane at the mid-channel height, full 3D velocimetry in channels of varying aspect ratio ($H/W$) would provide more comprehensive insights into the nature of the 3D twisting events we observed. Furthermore, direct measurements of the stress field during flow could provide deeper insights into the coupling between elastic stresses and flow structures (e.g., dead zones, 3D twisting events). Our experiments probed a wide range of shear rates, but they all fell on the stress plateau measured in the bulk shear rheology shown in Fig.~\ref{fig:rheo}(a); exploring a wider range of shear rates below and above this plateau will help further elucidate WLM flow dynamics across a broader range of conditions. 

Finally, while here we focused on a single semi-dilute, linear, entangled WLM solution, future studies could explore other formulations with different rheological properties (e.g., varying the degrees of shear-thinning and elasticity by tuning the molar ratio of surfactant to salt). Studies could also probe how variations in the micellar microstructure~\cite{chellamuthu_distinguishing_2008,oelschlaeger_linear--branched_2009,rogers_rheology_2014,gaudino_adding_2015} influence the flow behavior---which could potentially guide ways to use stimulus-responsive WLMs~\cite{shi_photoreversible_2013,chu_smart_2013} to achieve targeted local flow behaviors. For example, controllably forming stagnant dead zones could be a mechanism to redirect flow on demand, while conversely, prevention of dead zone formation could help promote fluid homogenization.\\

\textbf{Acknowledgements.} We acknowledge support from the Princeton Center for Complex Materials (PCCM), a National Science Foundation (NSF) Materials Research Science and Engineering Center (MRSEC; DMR-2011750, as well as the Camille Dreyfus Teacher-Scholar Program of the Camille and Henry Dreyfus Foundation. We also acknowledge the use of the Imaging and Analysis Center (IAC) operated by the Princeton Materials Institute at Princeton University. It is a pleasure to thank Howard Stone for useful feedback. \\

\textbf{Author contributions.} E.Y.C.: conceptualization, methodology, investigation, formal analysis, writing; S.S.D.: conceptualization, resources, supervision, writing.\\

\textbf{Data availability.}
Supplementary movies are available online at \href{https://github.com/emchen6639/WLMserpentine}{https://github.com/emchen6639/WLMserpentine}. \\

\section*{\label{app}Appendix: Geometric prediction of dead zone shape and size}
\begin{figure}
\centering
  \includegraphics{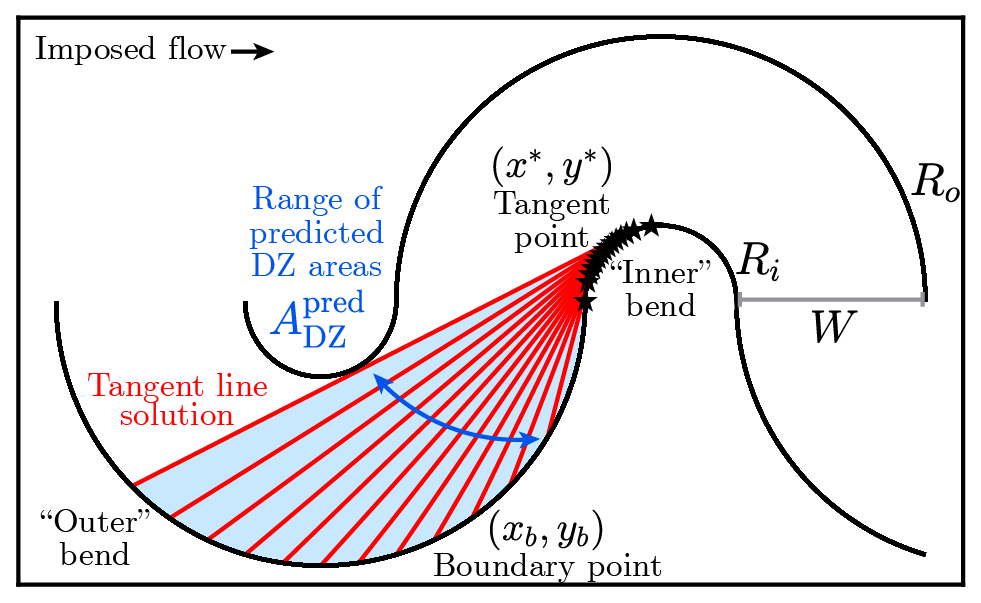}
  \caption{Schematic describing the geometric prediction of dead zone size and shape in the serpentine channels.}
  \label{fig:app}
\end{figure}

For one unit cell of the serpentine geometry, the upper boundary can be described as   
\begin{equation*}
y_{\mathrm{top}}(x)=
\begin{cases}
        \sqrt{R_o^2-(x-R_o)^2} &  0\leq x \leq2R_o\\
        -\sqrt{R_i^2-(x-2R_o-r_i)^2} &  2R_o\leq x \leq2R_i+2R_o,
    \end{cases}
\end{equation*}
where $x$ and $y$ represent the streamwise and transverse position coordinates. Similarly, the bottom boundary can be described as
\begin{equation*}
y_{\mathrm{bot}}(x)=
\begin{cases}
        \sqrt{R_i^2-(x-R_o)^2} &  W\leq x \leq W+2R_i\\
        -\sqrt{R_o^2-(x-2R_o-R_i)^2} &  W+2R_i \leq x \leq W+2R_i+2R_o.
    \end{cases}
\end{equation*}
Given a selected fixed point $(x_b,y_b)$ on the outer boundary of a half loop, as depicted in Fig.~\ref{fig:app}, we seek to determine the location on the following inner bend $(x^*,y^*)$ such that the line between the two points is tangent to the inner half-loop boundary, thereby minimizing local streamline curvature. The slope between the two points is $m=\frac{y^*-y_b}{x^*-y_b}$, which is equal to the derivative of inner curve parameterization evaluated at $(x^*,y^*)$: $m=-\frac{(x^*-R_o)}{\sqrt{R_i^2-(x^*-R_o)^2}}$. Further, the point $(x^*,y^*)$ must lie on both the inner curve equation and the tangent line equation. Accordingly, $(x^*,y^*)$ is straightforwardly determined from the solution of:
\begin{equation*}
    -\frac{(x^*-R_o)(x^*-x_b)}{\sqrt{R_i^2-(x^*-R_o)^2}} = \sqrt{R_i^2-(x^*-R_o)^2}-y_b.
\end{equation*} Note that only one of two solutions is physical in the context of our experiments. By varying the selected point $(x_b,y_b)$, we obtain predictions of the dead zone shape and size where the dead zone is bounded by the resulting tangent line and the outer channel boundary. Example solutions are shown by the red lines in Fig.~\ref{fig:app}. The maximum dead zone size is set by the physical constraints of the bounding geometry as conservation of mass requires that some finite region of the channel must contain non-zero flow velocity. The resulting range of theoretical dead zone size predictions $0 \leq A_\mathrm{DZ}^\mathrm{pred}\leq A_\mathrm{DZ}^\mathrm{max}$ is given by the shaded region in Fig.~\ref{fig:DZ}(c) and is in good agreement with the experimentally-measured values of dead zone sizes.

\end{document}